\documentclass[11pt]{cernart}
\usepackage[dvips]{graphics,rotating,epsfig}
\usepackage{amsmath}
\usepackage{times} 
\usepackage{xspace}
\usepackage{array}

\newcommand{\GeV}{\ensuremath{\mbox{GeV}}\xspace}

\newcommand{\GeVc}{\ensuremath{\mbox{GeV}/c}\xspace}

\newcommand{\cm}{\ensuremath{\mbox{cm}}\xspace}
\newcommand{\mm}{\ensuremath{\mbox{mm}}\xspace}
\newcommand{\micron}{\ensuremath{\mu \mbox{m}}\xspace}

\newcommand{\mrad}{\ensuremath{\mbox{mrad}}\xspace}

\begin{document}
\begin{titlepage}
\rightline {CERN-PH-EP/2007-xxx}
\rightline {21 August 2007}
\vglue 0.1cm
\renewcommand{\textfraction}{0.05}
\sloppy

\title{\Large \bf Associated Charm Production  in Neutrino-Nucleus Interactions }

\author{CHORUS Collaboration}
\maketitle

\begin{abstract}

In this paper a search for
associated charm production both in neutral and charged current
$\nu$-nucleus  interactions is presented.  The improvement of
 automatic scanning systems in the \mbox{CHORUS} experiment 
allows an efficient search to be performed in emulsion for short-lived particles. 
Hence a search for rare processes, like the associated charm production, becomes possible through the
observation of the double charm-decay topology
with a very low background. About 130,000 $\nu$ interactions located in the emulsion target have been
 analysed. Three events with two charm 
decays have been observed in the neutral-current  sample with an
 estimated background of 
0.18$\pm$0.05. 
The relative rate of the associated charm cross-section in deep inelastic 
  $\nu$ interactions,  
$\sigma(c\bar{c}\nu)$/$\sigma_\mathrm{NC}^\mathrm{DIS}=
(3.62^{+2.95}_{-2.42}(\mbox{stat})\pm 0.54(\mbox{syst}))\times 10^{-3}$
has been measured. One event with two charm decays has been observed in charged-current
 $\nu_\mu$ interactions with an estimated background of  0.18$\pm$0.06 and 
the upper limit on associated charm production in charged-current interactions at 90\% C.L. 
%in charged-current  $\nu_\mu$ interactions 
has been found to be $\sigma (c\bar{c} \mu^-)/\sigma_\mathrm{CC} < 9.69 \times 10^{-4}$.

\end{abstract}

\vspace{5.0cm}

%%%%%%%%%%%%%%%%%%%%%%%%%%%%%%%%%%%%%%%%%%%%%%%%%%%%%%%%%%%%%%%%%%%%%%%%%%%%%%%%

\newpage

\begin{center}   
{\Large {CHORUS Collaboration}}
\end{center}
\vspace{0.1cm}

\begin{Authlist}
A.~Kayis-Topaksu, G.~\"{O}neng\"ut

{\bf \c{C}ukurova University, Adana, Turkey}

R.~van Dantzig,  M.~de Jong, 
%O.~Melzer, 
R.G.C.~Oldeman$^1$%,
%E.~Pesen, J.L.~Visschers

{\bf NIKHEF, Amsterdam, The Netherlands}

M.~G\"uler, U.~K\"ose, 
%M.~Serin-Zeyrek, 
%R.~Sever, 
P.~Tolun
%, M.T.~Zeyrek

{\bf  METU, Ankara, Turkey}

M.G.~Catanesi, 
%M.~De Serio, M.~Ieva,
M.T.~Muciaccia
%, E.~Radicioni, S.~Simone 

{\bf Universit\`a di Bari and INFN, Bari, Italy}

%A.~B\"ulte, 
K.~Winter

{\bf  Humboldt Universit\"at, Berlin, Germany$^{2}$}

B.~Van de Vyver$^{3,4}$, P.~Vilain$^{5}$, G.~Wilquet$^{5}$

{\bf Inter-University Institute for High Energies (ULB-VUB) Brussels, Belgium}

B.~Saitta

{\bf Universit\`a di Cagliari and INFN, Cagliari, Italy}

E.~Di Capua

{\bf Universit\`a di Ferrara and INFN, Ferrara, Italy}

S.~Ogawa, H.~Shibuya

{\bf Toho University,  Funabashi, Japan}

%A.~Artamonov$^6$, M.~ Chizhov$^7$, M.~ Doucet$^{8}$, 
I.R.~Hristova$^6$, 
T.~Kawamura,
D.~ Kolev$^7$, H.~ Meinhard, 
J.~Panman, 
%I.M.~Papadopoulos, S. Ricciardi$^{11}$,
A.~Rozanov$^{8}$,
R.~Tsenov$^{7}$, J.W.E. Uiterwijk, P. Zucchelli$^{3,9}$

{\bf CERN, Geneva, Switzerland}

J.~Goldberg

  {\bf Technion, Haifa, Israel}

M.~Chikawa

 {\bf Kinki University, Higashiosaka, Japan}

%E.~Arik
%
% {\bf Bogazici University, Istanbul, Turkey}

J.S.~Song, C.S.~Yoon

{\bf Gyeongsang National University,  Jinju, Korea}

K.~Kodama, N.~Ushida

{\bf Aichi University of Education, Kariya, Japan}

S.~Aoki, T.~Hara

 {\bf Kobe University,  Kobe, Japan}

T.~Delbar,  D.~Favart, G.~Gr\'egoire, S.~ Kalinin, I.~ Makhlioueva

{\bf Universit\'e Catholique de Louvain, Louvain-la-Neuve, Belgium} 

A.~Artamonov, P.~Gorbunov$^{3}$, V.~Khovansky, V.~Shamanov, I.~Tsukerman

{\bf Institute for Theoretical and Experimental Physics, Moscow, Russian
Federation}

N.~Bruski, D.~Frekers

{\bf Westf\"alische Wilhelms-Universit\"at, M\"unster, Germany$^{2}$}

K.~Hoshino, 
J.~Kawada, 
M.~Komatsu,
M.~Miyanishi, 
M.~Nakamura, T.~Nakano, K.~Narita, K.~Niu, K.~Niwa, 
N.~Nonaka, O.~Sato, T.~Toshito

{\bf Nagoya University, Nagoya, Japan}

S.~Buontempo, A.G.~Cocco, N.~D'Ambrosio,
G.~De Lellis, G.~ De Rosa, F.~Di Capua, 
%A.~Ereditato, 
G.~Fiorillo, A.~Marotta,
%M.~Messina, 
P.~ Migliozzi, 
%C.~Pistillo, 
L.~Scotto Lavina, 
%M.~Sorrentino,
P.~ Strolin, V.~Tioukov

{\bf Universit\`a Federico II and INFN, Naples, Italy}

%K.~Nakamura, 
T.~Okusawa

{\bf Osaka City University, Osaka, Japan}

U.~Dore, P.F.~Loverre,
L.~Ludovici, 
%P.~ Righini,
G.~Rosa, R.~Santacesaria, A.~Satta, F.R.~Spada

{\bf Universit\`a La Sapienza and INFN, Rome, Italy}

E.~Barbuto, C.~Bozza, G.~Grella, G.~Romano, C.~Sirignano, S.~Sorrentino

{\bf  Universit\`a di Salerno and INFN, Salerno, Italy}

Y.~Sato, I.~Tezuka

{\bf Utsunomiya University,  Utsunomiya, Japan}

{\footnotesize
---------

\begin{flushleft}

%$^{1}$ {Now at Universit\`a La Sapienza, Rome, Italy.}
$^{1}$ Now at Universit\`a di Cagliari, Cagliari, Italy.%Now at University of Liverpool, Liverpool, UK
\newline
$^{2}$ {Supported by the German Bundesministerium f\"ur Bildung und Forschung under contract numbers 05 6BU11P and 05
7MS12P.}
\newline
$^{3}$ {Now at SpinX Technologies, Geneva, Switzerland.}
\newline
$^{4}$ {Fonds voor Wetenschappelijk Onderzoek, Belgium.}
\newline
$^{5}$ {Fonds National de la Recherche Scientifique, Belgium.}
\newline
%$^{6}$ {On leave of absence from ITEP, Moscow.}
%\newline
$^{6}$ {Now at DESY, Hamburg.}
%\newline
%$^{7}$ {On leave of absence from \c{C}ukurova University, Adana, Turkey.}
\newline
$^{7}$ {On leave of absence and at St. Kliment Ohridski University of Sofia, Bulgaria.}
\newline
%$^{8}$ {Now at University of Maryland, MD, USA.}
%\newline
%$^{11}$ {Now at Royal Holloway College, University of London, Egham, UK.}
%\newline
$^{8}$ {Now at CPPM CNRS-IN2P3, Marseille, France.}
\newline
$^{9}$ {On leave of absence from INFN, Ferrara, Italy.}
\end{flushleft}
}

\end{Authlist}

\end{titlepage}

\newpage

\section{Introduction}

\begin{figure}[h]
  \begin{center} \resizebox{0.7\textwidth}{!}{
    \includegraphics{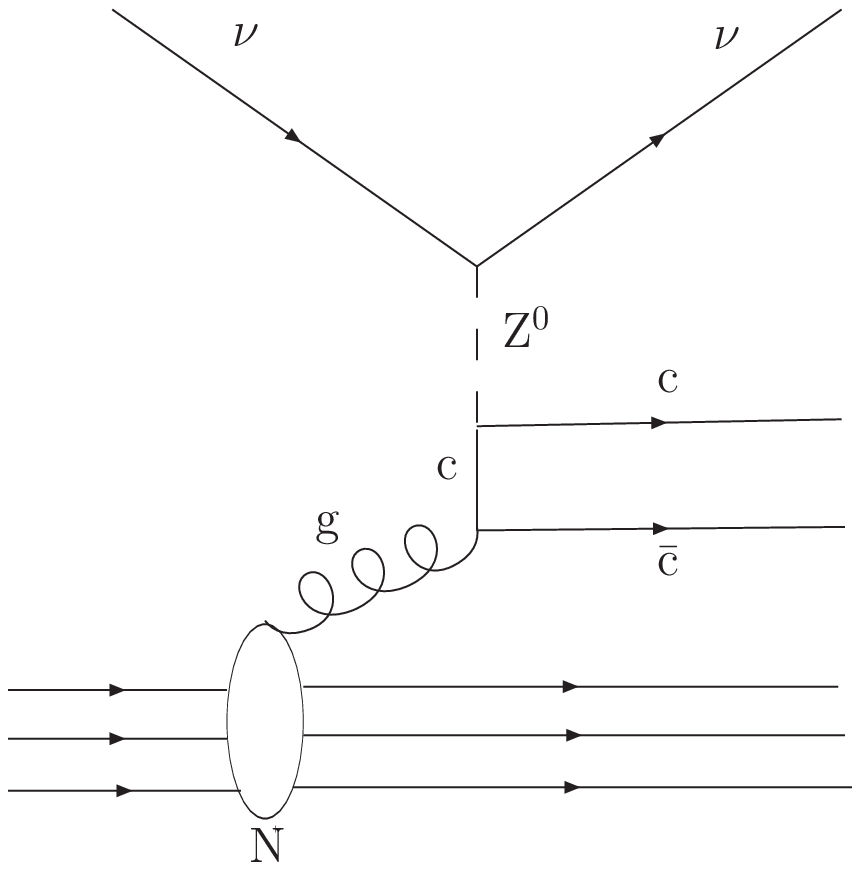}
\hspace{3.0cm}
    \includegraphics{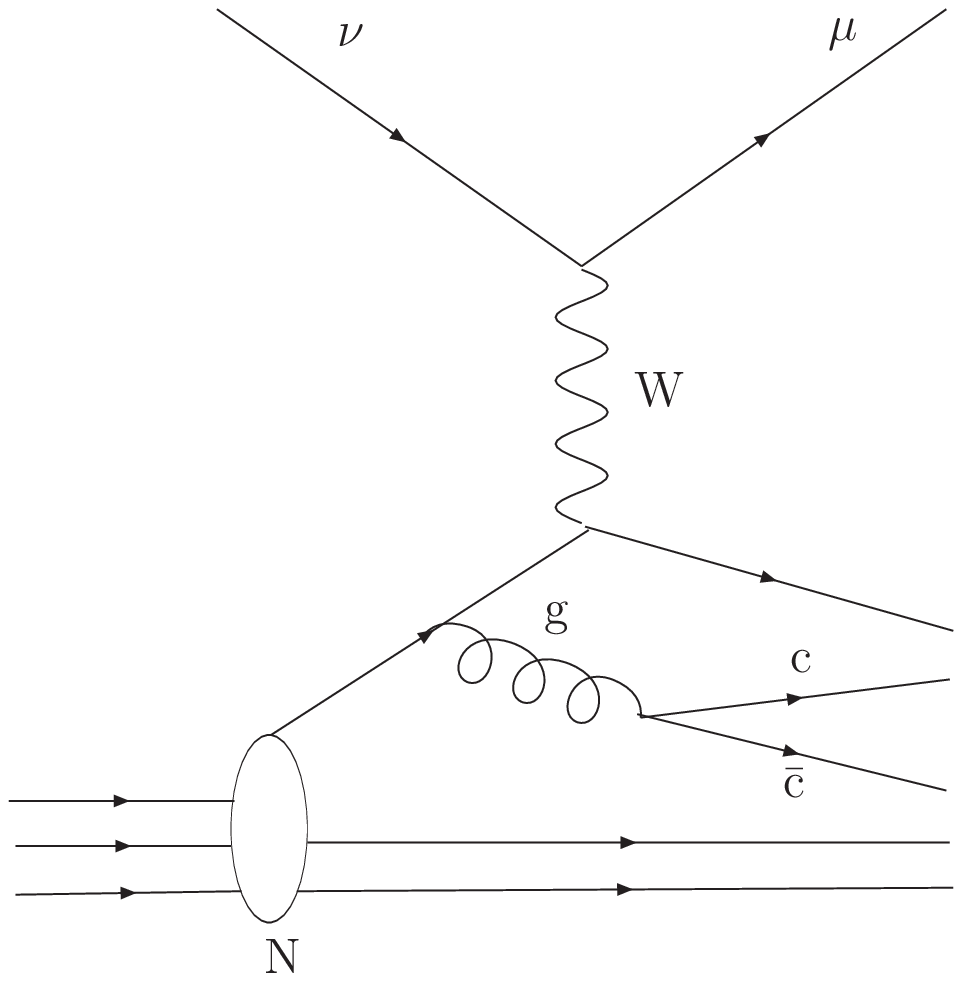} }\caption{Feynman diagrams for boson-gluon fusion(left) 
and gluon bremsstrahlung processes(right).}
\label{fig:feyn} \end{center}
\end{figure}

The production of two charmed particles in neutrino-nucleon scattering is a very rare process and therefore difficult
to measure. The charm pair originates from two different processes: one, the so-called boson-gluon fusion mechanism, is
possible only in neutral-current interactions, while the other, gluon bremsstrahlung, occurs both in neutral and
charged current (NC and CC)  interactions. The Feynman diagrams of both process are shown in Figure~\ref{fig:feyn}.

In early neutrino experiments at the CERN SPS and Fermilab,
measurements of production rates of prompt trimuons~\cite{benvenuti}
and like-sign dimuons (\cite{ben2},\cite{sakumoto} and \cite{land}) were found to
be higher than expected \cite{Hagiwara:1980nu}.  In the past the origin of these
events was assumed to be the associated charm production in
CC neutrino interactions with subsequent muonic
decay of the charmed particle(s):
\begin{equation}
\label{eq:proc}
\nu_{\mu} N \rightarrow \mu^{-} c \bar{c} X \end{equation} where the
charm-anticharm pair is produced by a gluon emitted via bremsstrahlung
from a light parton. More recently, an analysis by CHORUS~\cite{trimuon} suggested a significant contribution from
$\mu^{+}\mu^{-}$ decays of light vector mesons.
In a previous search, based on the visual observation of the charmed hadron decays, 
the CHORUS collaboration has published the first evidence for this process~\cite{ccbar} through the observation of one event.

For the NC process, the E531 experiment reported the
observation of one event with two neutral decay topology. Using this
event and correcting for the efficiency, a relative rate was quoted as 
$\frac{\sigma (\nu_{\mu} \rightarrow c\bar{c} \nu_{\mu} )}{\sigma
(\nu_{\mu} \rightarrow \nu_{\mu})} = (1.3_{-1.1}^{+3.1})\times
10^{-3}$~\cite{e531} at the average neutrino beam energy of 22~\GeV. 
No background estimation was given.

A measurement of this rate was recently made by the NuTeV experiment
~\cite{nutev} through the detection of events with wrong sign
muon final states. The analysis technique consists in comparing
the visible inelasticity, $y_\mathrm{vis}=E_\mathrm{HAD}/\left( E_\mathrm{HAD}+E_{\mu
}\right)$, measured in the $\nu _{\mu } $ and $\bar{\nu}_{\mu }$ data samples where the leading
muon had the wrong sign of charge, to a Monte Carlo simulation
containing all known conventional wrong sign muon sources and a possible NC charm
signal. The NC charm signal peaks at large values of $y_\mathrm{vis}$ because
the  muon from the decay of a heavy flavor hadron is usually much less
energetic than the hadron shower produced in the NC interaction.
After consideration of all background sources, a production cross
section $\frac{\sigma (\nu_{\mu} \rightarrow c\bar{c} \nu_{\mu} )}{\sigma
(\nu_{\mu} \rightarrow \nu_{\mu})} = (6.4_{-4.6}^{+5.5})\times 10^{-3}$
~\cite{nutev} at the average neutrino beam energy of 154 GeV was obtained.
More details both on
the theoretical and experimental aspects of associated charm production may be
found in Ref.~\cite{giovanni, migliozzi} and references in them.

All the results, available to date, with the exception those from 
 E531, were extracted from
multi-muon samples with a large background of non prompt muons.
In the CHORUS experiment a different search for associated charm
production in NC and CC $\nu_{\mu}$ interactions was carried out in nuclear
emulsion, is practically background free and is presented in this paper.
The search is based on the visual
observation of the primary vertex and the two decay vertices, taking
advantage of the sub-micrometric position resolution of the nuclear  emulsion.  
We report the observation of four 
events with two charm decays: three in NC and one in CC   interactions. 
Based on the observed events, taking care of the estimated background and correcting for the detection efficiencies, 
we obtain a production rate of associated charm in NC
interactions and an upper limit for the CC process.

\section{The experimental apparatus}   

The CHORUS detector is a hybrid setup that combines a nuclear emulsion
target with various electronic detectors, as described in Ref.~\cite{chorus}.

%The use of nuclear emulsions as target for neutrino interactions
%allows for the three-dimensional reconstruction of short lived
%particles like the $\tau$ lepton and charmed hadrons to be performed. 
A three-dimensional reconstruction of the decays of short-lived
particles such as the $\tau$ lepton and charmed hadrons can be performed
in nuclear emulsions.
The emulsion
target is segmented into four stacks with an overall mass of 770 kg.
Each of the stacks consists of eight modules of 36 plates of 36~\cm $\times$
72~\cm size.  Each plate consists of a 90 \micron plastic support
coated on both sides with a 350 \micron emulsion
layer~\cite{emulsion}. Each stack is followed by three
emulsion sheets, having a 90  \micron emulsion layer on both sides of an
800 \micron thick plastic base, acting as an interface to a set of scintillating~ fibre~
tracker planes.  The accuracy of the fibre tracker predictions is about
150 $\mu$m in track position and 2 \mrad in angle. The interface
sheets and the fibre trackers provide accurate 
particle trajectory
predictions in the emulsion stack in order to locate the interaction vertex.
The information of the electronic detectors has been used to define two data sets, the $1\mu$ and  $0\mu$ samples. The events belonging to the  
 $1\mu$ sample contain one reconstructed muon of negative charge. The muon identification and 
reconstruction  is based on the muon spectrometer response.
The $0\mu$ sample contains events where no muon is found. It mainly consists of NC $\nu_\mu$ interactions with
a contamination of mis-identified CC $\nu_\mu$  interactions and
interactions generated by neutrinos other than $\nu_\mu$. 

The emulsion scanning is performed by fully automated microscopes
equipped with CCD cameras and a read out system called 
``Ultra Track Selector''
~\cite{nakano}. In order to recognize the track segments in 
the emulsion, a series of tomographic images are taken by moving the focus at
different depths in the emulsion thickness.  The digitized images are
shifted according to the predicted track angle and then added.  The
presence of aligned grains forming a track is detected as a local peak
above the gray level of the summed image.  The track finding efficiency
of the track selector is higher than 98\% for track angles up to
400~\mrad with respect to normal incidence. 

The electronic detectors downstream of the emulsion target include a
hadron spectrometer which measures the bending of charged particles in
an air-core magnet, a calorimeter where the energy and direction of
showers are measured and a muon spectrometer which measures the charge
and momentum of muons. In this analysis, the momentum of particles other
than muons in the  
candidate events were determined by measuring their multiple scattering in the emulsion target.

\section{Event samples and  selection of two charm topologies}

The West Area Neutrino Facility (WANF) at CERN provided a wide band beam of 27
\GeV average energy consisting mainly of $\nu_\mu$, with a contamination of 5.1$\%$
$\bar{\nu}_{\mu}$, 0.8$\%$ ${\nu}_{e}$ and 0.2$\%$ $\bar{\nu}_{e}$.  During the four 
years of CHORUS operation
the emulsion target has been exposed to the beam for an integrated
intensity which corresponds to $5.06 \times 10^{19}$ protons on
target.  The data from the electronic detectors have been analyzed and
the set of events possibly originating in the emulsion stacks has
been identified.

In total about 130,000 $\nu$ events have been located in emulsion with
a procedure described in Ref.~\cite{oscil}. In order to identify a charm
decay, a volume of 1.5~\mm$\times$1.5~\mm$\times$6.3~\mm around the located
vertex position is scanned.  The parameters of
all track segments with angles below 400 \mrad found in this volume are
stored in a database. Typically, five thousand track segments are
recorded per event. This procedure is called ``Netscan''~\cite{netscan}.
A detailed description of the procedure and of the algorithms used 
to reconstruct 
the event topology are given in Ref.~\cite{murat, netscan2}.

Track segments collected in the Netscan fiducial volume are analysed 
off-line to reconstruct the complete event topology: 
segments from different plates are combined into tracks; low-energy 
tracks are filtered out as well as track originating 
from outside the volume (mainly passing-through beam muons).

Possible vertices are then defined, requiring that the minimum distance 
between tracks attached to a common vertex is compatible with zero within 
errors.  If several vertices are reconstructed, the primary vertex is defined, 
for the $1\mu$ sample, as the one to which the muon track is attached. For 
the $0\mu$  sample, the primary vertex is chosen as the most upstream 
one with at least one track matching one of the tracks reconstructed in 
the electronic detectors.
Events for which the emulsion tracks do not converge to a single common 
vertex potentially contain a short-lived particle decay topology. 
They were selected for further inspection if one of the following selection 
criteria is satisfied ~\cite{murat, netscan2} :
\begin{itemize}
\item Two or more multi-tracks vertices are reconstructed.
\item At least one track of the primary vertex is not reconstructed up to 
the edge of the Netscan volume
\item A track originating in the same plate as the primary vertex has an 
impact parameter {\it IP} relative to it satisfying the two conditions:
\begin{itemize}
\item {\it IP} $>$ $\sqrt{ (5.0^{2}+(2 \mbox{d}x \times \sigma)^{2} ) }$~\micron 
where 
$\sigma=\sqrt{ (0.003^{2}+(0.0194 \times \tan\theta)^{2}) }$ is a
parameterization of the angular error and d$x$ is the distance of the
vertex from the most upstream daughter track segment.
\item  {\it IP} $<$ 750 \micron,  to reject spurious tracks not related to the 
neutrino interaction.
\end{itemize}
\end{itemize}

After selection, the samples of 26,621 $0\mu$ and 99,245 $1\mu$ events reduce 
to respectively 717 and 2816 events.
These are then visually inspected to confirm the decay topology. A
secondary vertex is accepted as decay if 
the number of prongs is consistent with charge conservation and no evidence 
for hadronic interaction, e.g. Auger electron, nuclear recoil or break up, is 
observed.
Decays into a single charged particle (so called kink) are accepted only if 
the angle between the parent and daughter tracks is greater than 50 \mrad and 
the length of the parent track is greater than 25 \micron .

The observable decay topologies are classified according to their 
numbers of charged decay products into C1, C3 or C5 for the charged and V2, V4 
or V6 for neutral decaying particles.
The result of the 
visual inspection is given in Table~\ref{tab:0mu} and
Table~\ref{tab:1mu} for the $0\mu$ and  the $1\mu$ samples respectively.  
In both the $0\mu$ and the $1\mu$ samples, three events have been observed 
with two charm-decay topology.

The rejected sample consists mainly of low momentum tracks for which the 
multiple scattering leads to a fake impact parameter or to a failure to 
reconstruct the track as traversing the plate. The 
remaining events in the sample consist of hadronic interactions, gamma
 conversions and overlay secondary  vertices, 
reconstructed using one or more background tracks or of vertices with a
 parent track not connected to the primary vertex.  

After topological confirmation of the decays, further kinematical selections
are applied in order to eliminate the background 
coming from strange particle decays and from hadronic interactions. 
In case of a C1 topology, the requirement $p_\mathrm{T}>0.25$~\GeVc on the 
transverse 
momentum of the daughter particle with respect to the parent direction 
is applied. This cut also reduces significantly the background caused 
by ``white kinks'', i.e. single prong hadronic interaction without 
emission of any heavily ionizing particle, Auger electron or other 
evidence for nuclear recoil or break up.
The selection criterion applied to V2 decays is $\phi >$10~\mrad, where 
the angle $\phi$ measures the acoplanarity between the parent direction and 
the plane formed by the charged decay products. This cut eliminates mainly 
the two-body decays of neutral strange particles.
The effects of these kinematical cuts are included in the determination of the
Netscan efficiencies.

Four events satisfied the above selection criteria. 
Their general features are listed in Tables~\ref{tab:gfea} and \ref{tab:fea}.
%They are described in the next section.
\begin{table} [t]
\begin{center}
\caption {Results of the visual inspection in the selected $0\mu$ sample.}
\label{tab:0mu}
\vspace{0.5\baselineskip}
\begin{tabular}{cc|cc}
\multicolumn{2}{c|}{\bf Charm candidates}&
\multicolumn{2}{c}{\bf Rejected events} \\ \hline
{\bf Topology} & {\bf Events} & {\bf Category} & {\bf Events} \\
\hline
 V2&   145 & Low momentum &  72 \\
 V4&   43  & Traversing tracks  &  108 \\
 C1& 80    & $h^{\pm}$ int. & 110 \\
 C3& 75    & $h^0$ int. & 4 \\
 C5& 4     & $\gamma$- conversion & 36 \\
 V2+V2& 1  & Overlay secondary vertices & 37 \\
 C1+V2& 1  &~&~\\
 C3+V4& 1  &~&~\\
\hline
Total &   350 & ~& 367 \\
\end{tabular}
\end{center}
\end{table}
\begin{table} [h]
\begin{center}
\caption {Results of the visual inspection in the selected 1$\mu$ sample.}
\label{tab:1mu}
\vspace{0.5\baselineskip}
\begin{tabular}{cc|cc}
\multicolumn{2}{c|}{\bf Charm candidates}&
\multicolumn{2}{c}{\bf Rejected events} \\ \hline
{\bf Topology} & {\bf Events} & {\bf Category} & {\bf Events} \\
\hline
V2 & 841 & Low momentum & 140 \\
V4 & 230 & Traversing tracks  & 149\\
V6 & 3   & $h^{\pm}$ int. & 258 \\
C1 & 462 & $h^0$ int. & 69  \\
C3 & 501 & $\gamma$- conversion & 101 \\
C5 & 23  & Overlay secondary vertices& 36\\
C1+C3 & 1 & & \\
C1+V2 & 1 & & \\
V2+V4 & 1 & & \\
\hline
Total & 2063 &  & 753  \\
\end{tabular}
\end{center}
\end{table}

\section{The candidate events in the NC sample}

In the NC sample three candidates for associated charm production satisfied the  above
selection criteria. A description of each event is given below.

\begin{figure}[t]
  \begin{center} \resizebox{0.7\textwidth}{!}{
    \includegraphics{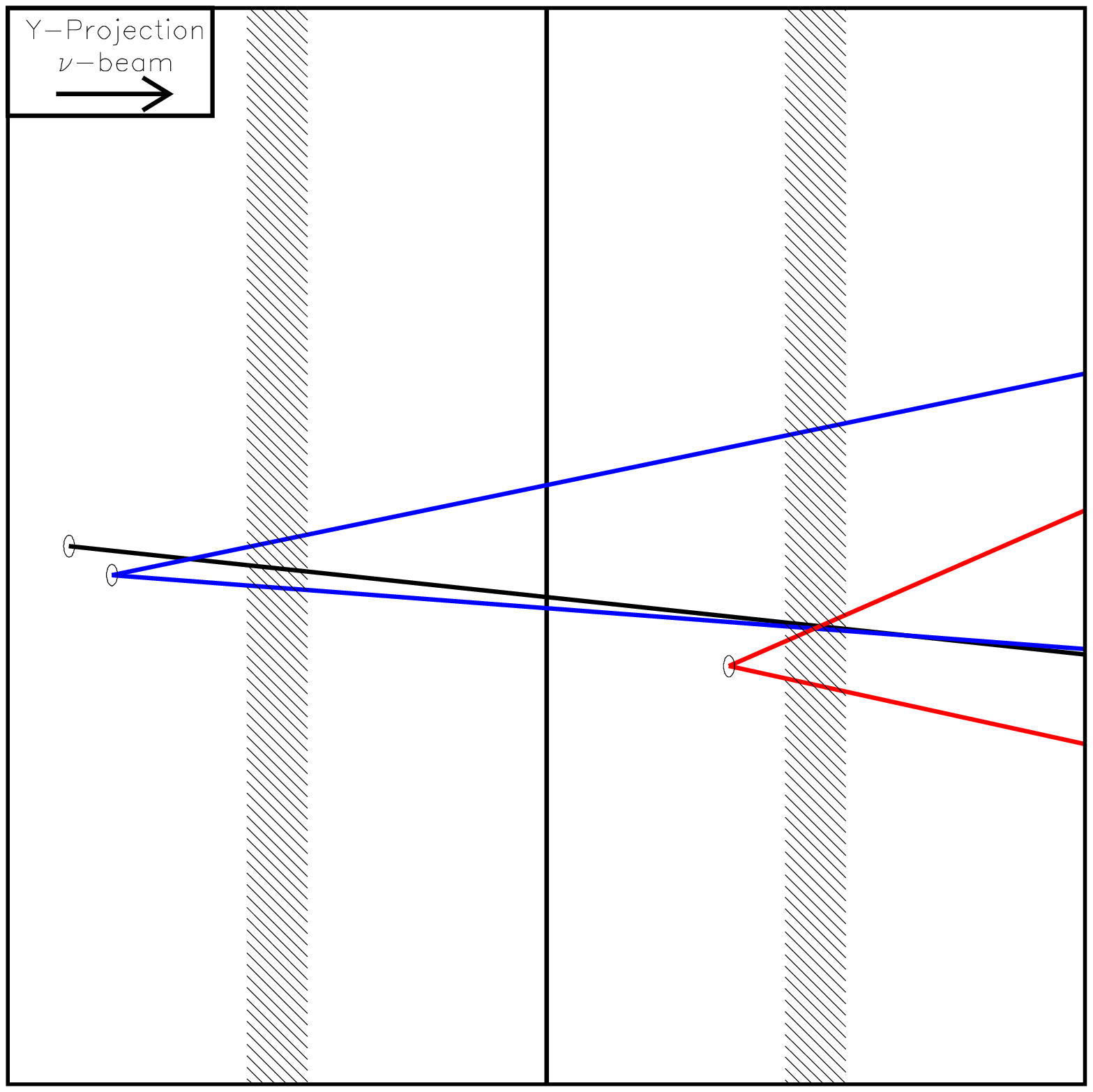} 
\hspace{3.0cm}
    \includegraphics{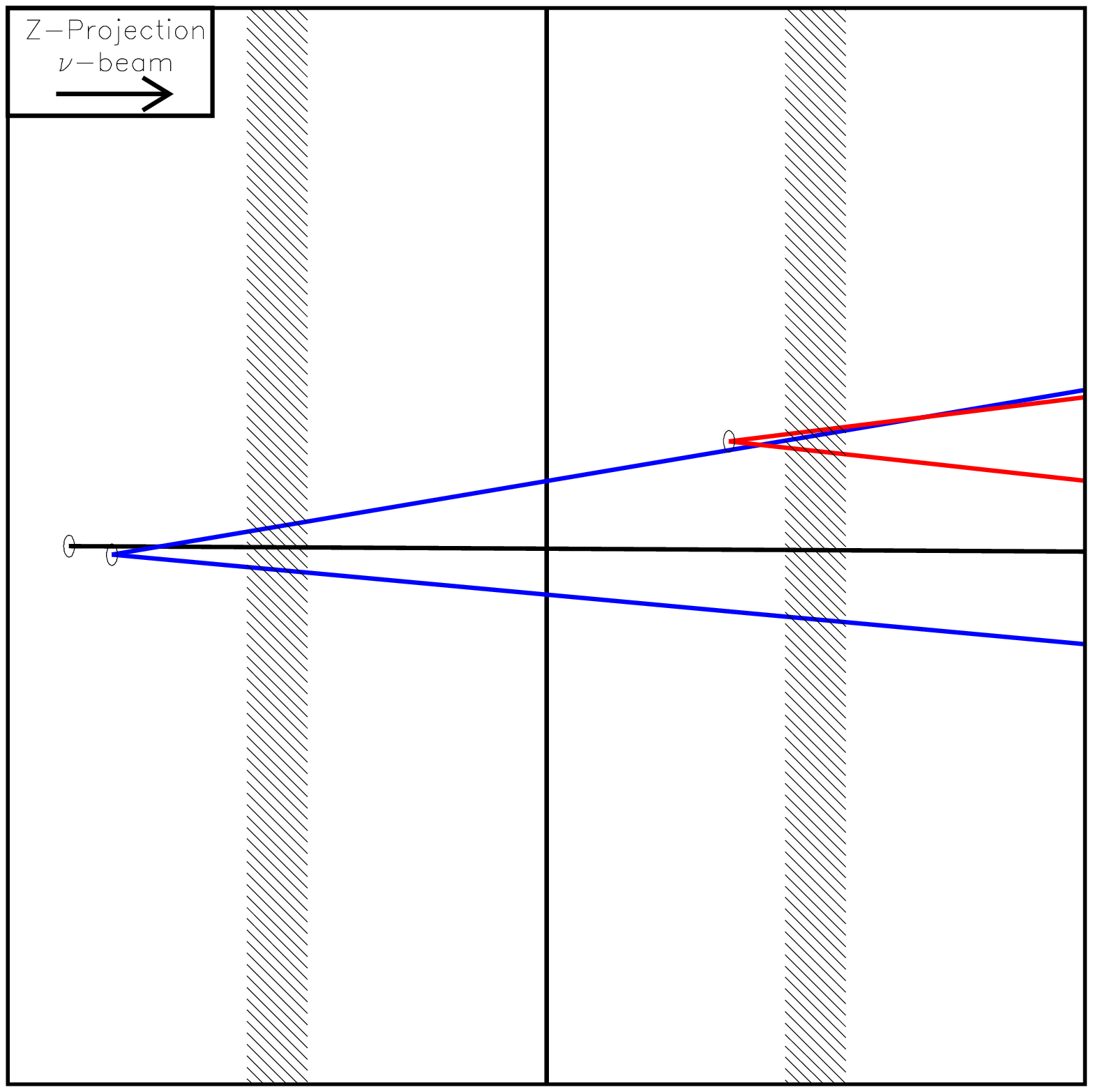} }\caption{Sketch of the NC event,
   (8132-12312) with a V2+V2 topology (the shaded area is the 90 \micron
   plastic base between the two 350 \micron emulsion layers).}  
\label{fig:8133} \end{center}
\end{figure}

\begin{figure}[t]
  \begin{center} \resizebox{0.9\textwidth}{!}{
    \includegraphics{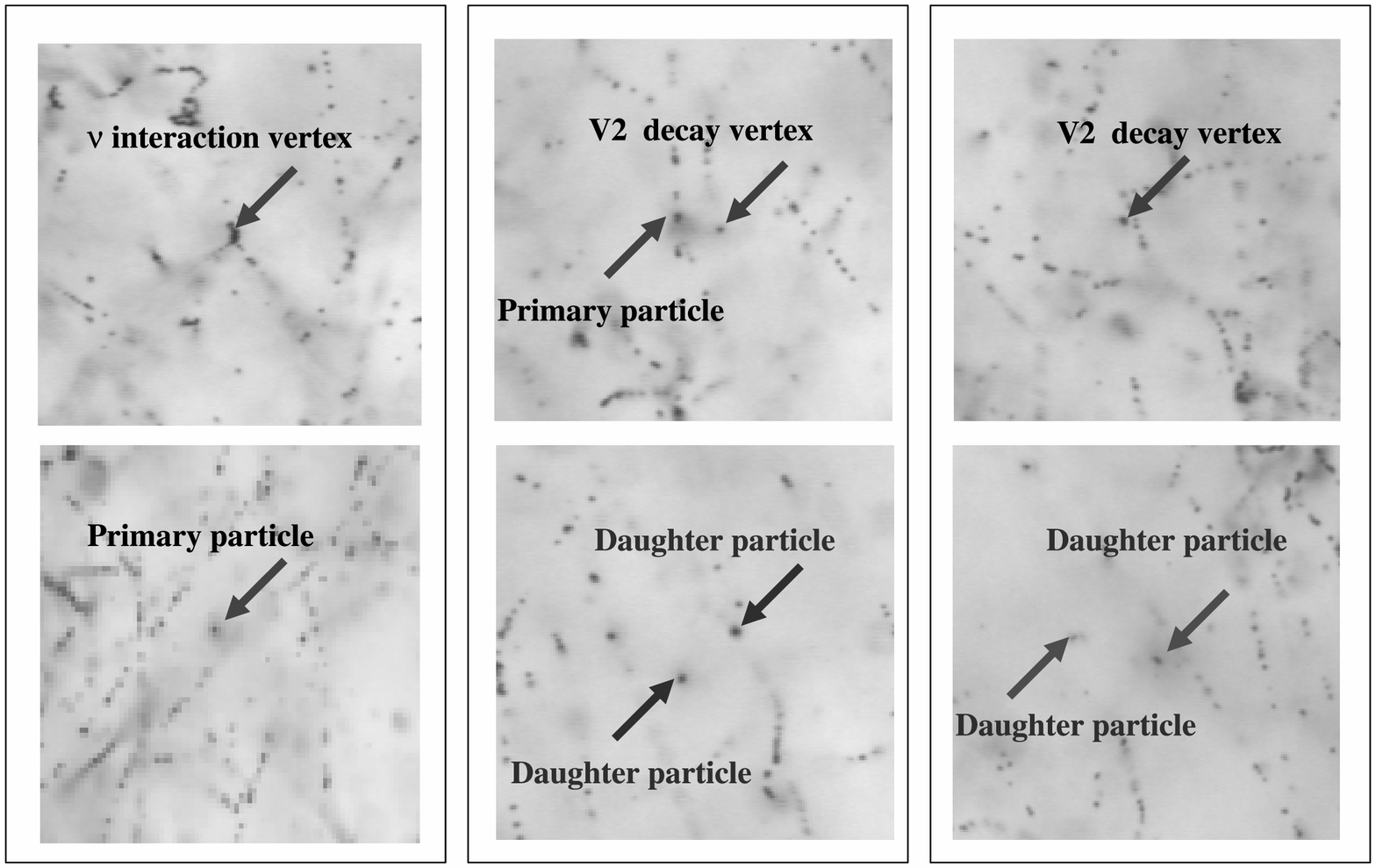} }\caption{Microscope views 
of Event 8132-12312  taken in the emulsion. Since the neutrino beam is perpendicular to the emulsion target,
the trajectory of a charged particle appears as a  'dot'. The  top-left  picture shows the neutrino
interaction vertex. A
charged particle which is visible several microns downstream of the vertex emerges from the neutrino  vertex. 
The top-center picture
 shows the decay vertex  of  a neutral charmed particle decaying into two
charged particles (V2)  63 \micron downstream of the primary vertex. The trajectory of the daughter
particles is visible at bottom-center picture. A second V2 decay topology is  found 
977 \micron downstream of the  primary vertex as shown in the top-right picture. In the bottom-right
picture the trajectory of daughter particles from the second neutral decay  becomes visible after
several microns downstream of  the decay vertex.}
\label{fig:v2v2} \end{center}
\end{figure}

Event 8132-12312 with a V2+V2 decay topology is sketched in Figure~\ref{fig:8133} and 
microscope view of the event in the emulsion at relevant positions is shown in Figure~\ref{fig:v2v2}.
Only one minimum-ionizing 
("shower") particle is emitted from the primary vertex and is identified as a hadron in the electronic detector.
 In the same emulsion plate, a neutral particle decays into two
charged particles (V2) 63 \micron downstream of the primary vertex. Both
daughter tracks are reconstructed in the electronic detector.  
The acoplanarity of parent and daughter tracks, 
$\phi$ = 24.2~\mrad $\pm$ 5.3~\mrad rules out a two body decay.   
Moreover, the lower limit of the mass $M_\mathrm{min}$ of the
parent particle is measured to be  0.80 \GeV (at 90\% C.L.).   
In one plate downstream, a second V2 decay has been located, 977 \micron from the 
primary vertex. The acoplanarity angle $\phi$ and the 
minimum mass $M_\mathrm{min}$  are  measured to be 
36.1~\mrad $\pm$ 0.3~\mrad and  0.95 \GeV (at 90\% C.L.) respectively. 
\begin{table} [h]
\begin{center}
\caption {General features of the associated charm candidates. Ns(Ns$^\star$) and Nh stand
for number of shower (reconstructed in the electronic detector)
and heavily ionizing tracks at the neutrino interaction vertex.
$E_{sh}$ is the  hadronic shower energy, $p_{\mu}$ the muon momentum
and $\theta_{\mu}$(y,z) the angle between the muon
and the beam directions, respectively in the horizontal and vertical planes. No correction was made
for the energy of the two neutrinos from the charm decays.}
\vspace{0.5\baselineskip}
\begin{tabular}{ccccccc} 
~& {\bf Event Id} & {\bf Ns~(Ns$^\star$)} & {\bf Nh} & ${\bf E_{sh}}${\bf(\GeV)}& ${\bf p_{\mu}}${\bf (\GeVc)}& 
${\bf \theta_{\mu}}${\bf(y,z)(rad)} \\
\hline
~&8132-12312 &1(1)& 2 & 38.2&~&~    \\
%\hline
NC& 7692-5575 &2(1)& 0 & 47.1&~&~    \\
%\hline
~&7739-3952 &6(1)& 1 & 45.9&~&~    \\
\hline
CC&7904-4944 &4(2)& 0 & 42.5&17.6&(-0.164,0.080)    \\
\end{tabular}
\label{tab:gfea}
\end{center}
\end{table}
\vspace{0.5\baselineskip}

\begin{table} [h]
\begin{center}
\caption {General features of the charged daughter particles in the associated charm 
candidates. 
$M_\mathrm{min}$  stands for  
minimum mass at 90 $\%$ C.L. of the parent particle.  $\theta$(y,z) is the angle between the charged daughter 
particle and the beam directions, respectively in the horizontal and vertical planes,
$p$ is the momentum of the  charged daughter particle measured using  multiple coulomb 
scattering. (The momentum measurement was not applied to Event 7739-3952 and failed for 
two daughters of V4 decay in Event 7739-3952.)}
\vspace{0.5\baselineskip}
\begin{tabular}{c|ccccc}
~& {\bf Event Id} & {\bf Topology} & $\bf{M_{min}(\GeV)}$ & {\bf{$\theta$}(y,z)(rad)} & \bf{p (\GeV/c)}\\
\hline
NC&8132-12312 &V2& 0.80& (0.049  -0.026) &4.66    \\
~&~&~& ~& (-0.023,0.040) &1.39    \\
~&~ &V2& 0.95& (-0.165,0.045) &2.72    \\
~&~ &~& ~& (0.034,0.105) &3.33    \\ 

~& 7692-5575 & V2 & 0.68& (0.026,-0.056)&4.11\\
~& ~& ~ & ~& (0.082,0.032)&3.81\\
~& ~& C1 & ~& (-0.136,0.291)&1.82\\ 
~&7739-3952 &V4& ~ & (0.129,-0.072)&~   \\
~&~&~& ~ & (-0.005,-0.064)&~   \\
~&~&~& ~ & (0.083,0.042)&~   \\
~&~&~& ~ & (0.075,0.336)&~   \\
~&~&C3& ~ & (0.011,-0.038)&~   \\
~&~&~& ~ & (-0.071,0.238)&~    \\   
~&~&~& ~ & (-0.294,-0.376)&~    \\   
\hline
CC&7904-4944 &V4& ~ &(0.007,-0.087) &1.92    \\
~&~&~& ~ &(-0.045,-0.001) &2.32    \\ 
~&~&~& ~ &(0.269,0.305) &~    \\ 
~&~&~& ~ &(0.101,-0.169)&~    \\
~&~&V2& 0.81 &(0.029,-0.053) &4.70    \\ 
~&~&~& ~ &(0.001,0.289) &0.67    \\ 
\end{tabular}
\label{tab:fea}
\end{center}
\end{table}

Event 7692-5575 (with a C1+V2 decay topology): At the primary vertex there are
two shower tracks, one of which decaying after 163 \micron into 
a charged particle (C1) with a 0.179 rad $\it{kink}$ angle. The daughter 
momentum was measured by multiple scattering to be larger than 
1.82~\GeVc (at 90\% C.L.). Its transverse momentum $p_\mathrm{T}$ 
with respect to its parent is larger than 0.32~\GeVc (at 90\% C.L.), 
which rules out the hypothesis of a strange particle decay.
In the same emulsion
plate, 224 \micron from the primary vertex, a second decay
with a V2 topology has been observed; the decay vertex occurs however in 
the plastic base.  The
acoplanarity angle is 14.9~\mrad $\pm$ 1.5~\mrad and 
$M_\mathrm{min}=$ 0.68 \GeV  (at 90\% C.L.) 
These values are inconsistent with a strange-particle decay hypothesis.
%\begin{figure}[t]
%  \begin{center} \resizebox{0.7\textwidth}{!}{
%    \includegraphics{7692_2y.eps}
%\hspace{3.0cm}
%    \includegraphics{7692_2z.eps} }\caption{Sketch of NC event
%   (7692-5575) with a C1+V2 topology (the shaded area is the 90 \micron 
%   plastic base between the two 350 \micron emulsion layers).}
%\label{fig:7692} \end{center}
%\end{figure}

Event 7739-3952 (with a C3+V4 decay topology): 
%is observed in nuclear emulsion.
%Figure~\ref{fig:napcand}.  
The primary vertex has six shower tracks, 
none of them identified as a muon.
One of the shower tracks decays after 426 \micron into three charged 
particles while the V4  decay vertex is 884 \micron from the primary 
vertex.

%\begin{figure}[b]
%  \begin{center} \resizebox{0.7\textwidth}{!}{
%    \includegraphics{7739_2y.eps}
%\hspace{3.0cm}
%     \includegraphics{7739_2z.eps} }\caption{Sketch of the event
%   (7739-3952) with a C3+V4 topology (the shaded area is the 90 \micron 
%   plastic base between the two 350 \micron emulsion layers).}
% \label{fig:napcand} \end{center}
%\end{figure}

\section{The candidate event in the CC sample }

\begin{figure}[ht]
  \begin{center} \resizebox{0.7\textwidth}{!}{
    \includegraphics{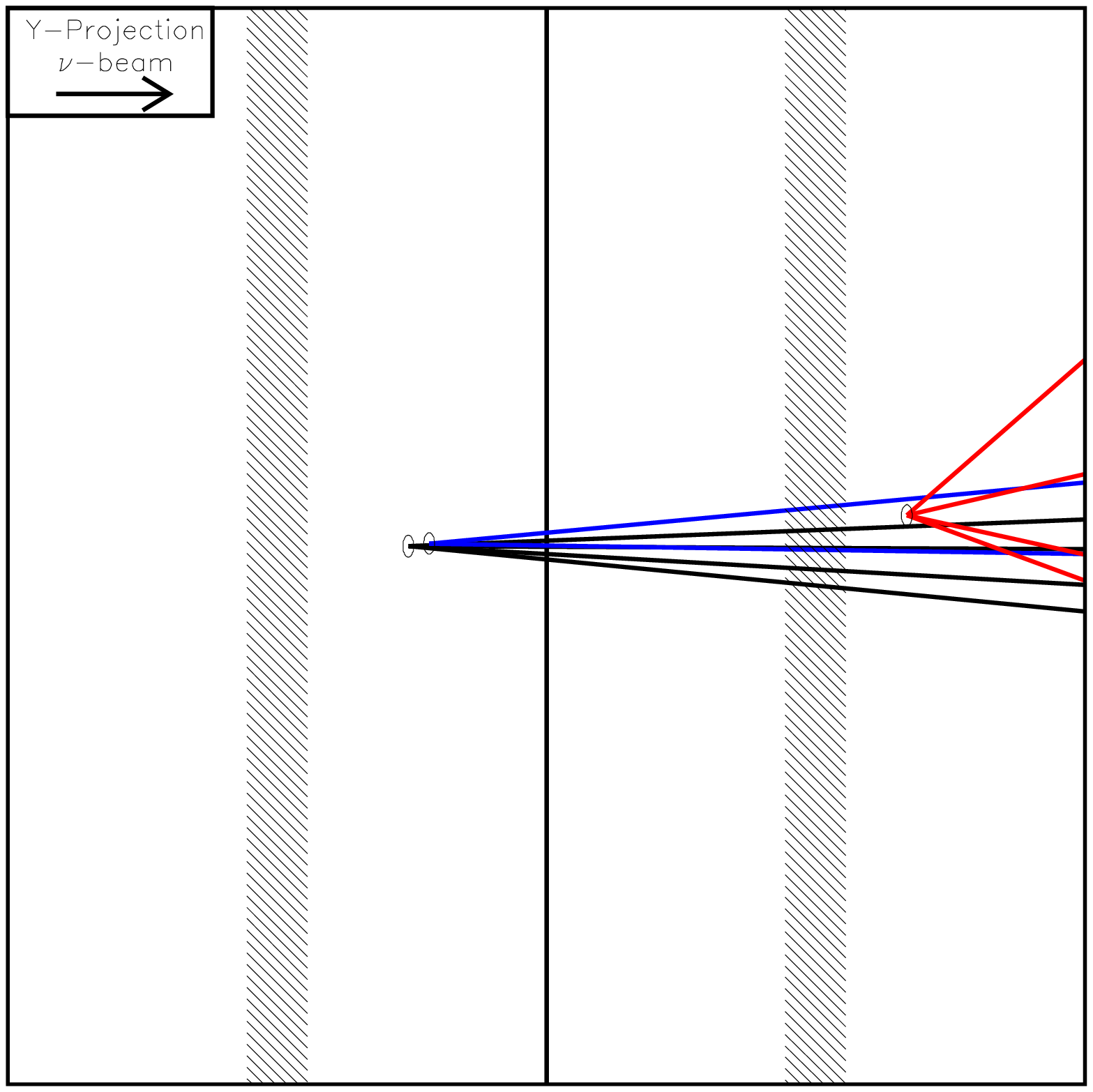}
\hspace{3.0cm}  
     \includegraphics{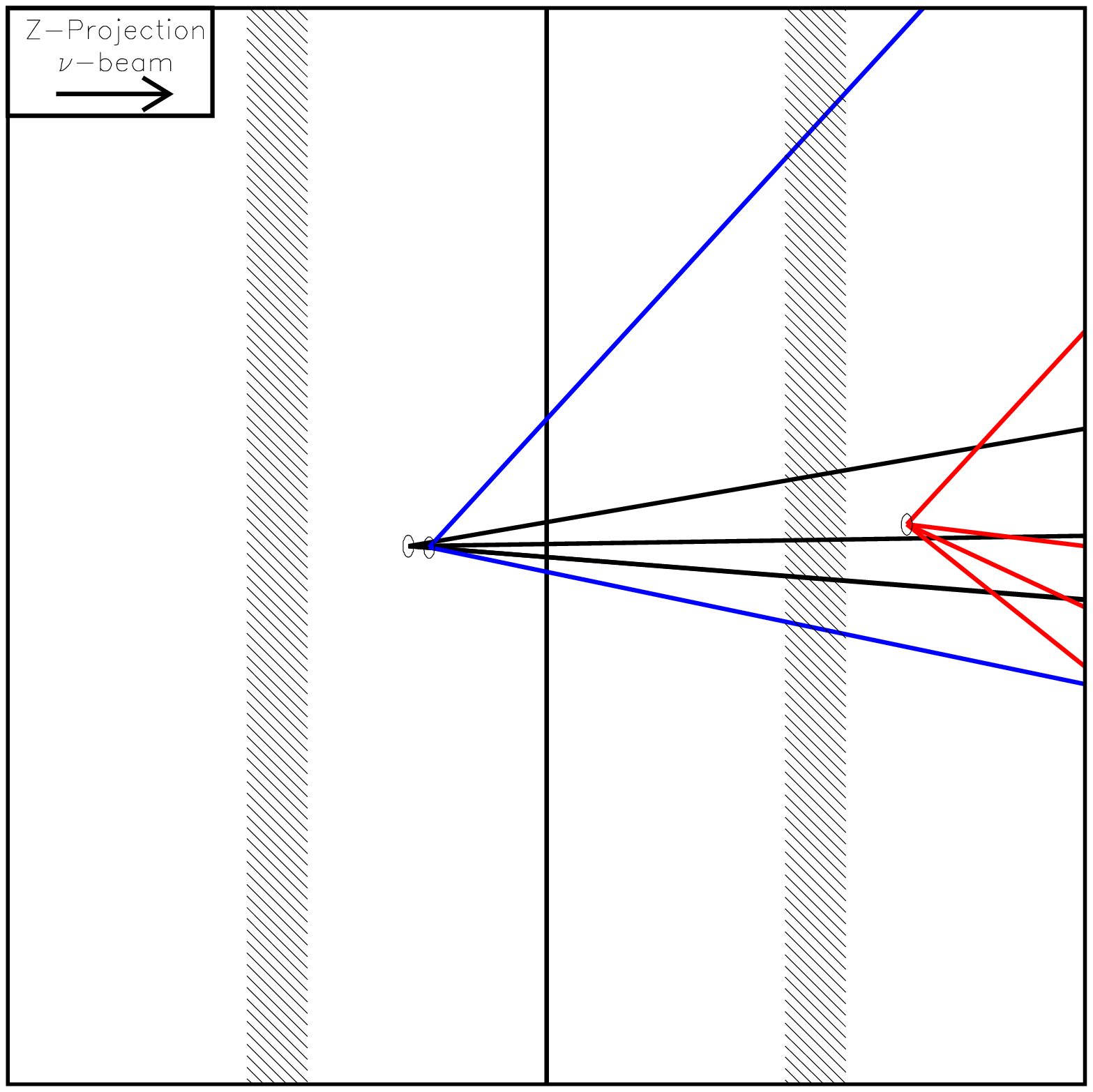} }\caption{Sketch of CC event
   (7904-4944) with a V2+V4 topology (the shaded area is the 90 \micron plastic base between the two 350 \micron emulsion layers).}
 \label{fig:7904} \end{center}
\end{figure}

Out of the three double charm candidates of Table 2, two are rejected on the 
basis of the ${p_\mathrm{T}}$ cut at their C1 vertex.
The remaining event 7904-4944 is shown in Figure~\ref{fig:7904}.
The primary vertex is located 
200~\micron from the plate downstream surface.  
There are four primary
shower tracks; one of them is identified as a negative muon in the
spectrometer. 
In the same plate, 58~\micron downstream of the primary vertex, a
neutral particle decays into a V2 topology. 
Both daughter tracks are reconstructed in the electronic detector. 
The acoplanarity angle $\phi$ is 12.8~\mrad $\pm$ 1.5~\mrad, and the minimum 
mass of the parent
particle, $M_\mathrm{min}=$~0.81 \GeV at 90\% C.L. 
A second neutral particle decays into a V4 topology with a
flight length of 761~\micron from the primary vertex. 
Two of four daughter tracks are
reconstructed in the electronic detector.  

In Ref.~\cite{ccbar}, we have reported the observation of one event in
CC $\nu_\mu$ interactions based on a subset of the data (about 50\% of the present
sample). This search was based on different selection criteria and was
performed before the development of the Netscan technique: charged
particles at the primary vertex were followed down manually to find
decay vertices. This event has not been retrieved in the present analysis because the 
emission angle of the charm daughter is out of the angular acceptance of the Netscan procedure.

\section{Evaluation of the reconstruction efficiency}

Large samples of neutrino interactions were generated with a neutrino
energy distribution according to the 
beam spectrum using the HERWIG event generator~\cite{herwig65}, with the 
leading-order parton distribution functions of MRST~\cite{MRST98LO}.
This generator
produces associated charm-production events in both neutral-current and
charged-current interactions. 
%via the two processes. 
Efficiencies and backgrounds were
evaluated with a simulation of the detector based on
GEANT3~\cite{geant}.

The simulated response of the electronic detectors is processed
through the same reconstruction program as the one used for the data. 
The tracks
in emulsion and the performance of the UTS are also simulated, in
order to evaluate the efficiency of the scanning procedure.

In order to evaluate the Netscan procedure efficiency, one needs to
reproduce realistic track densities.  This was achieved
by merging the emulsion data of the simulated events with real Netscan
data which do not have a reconstructed vertex but contain tracks which
stop or pass through the Netscan fiducial volume, representing the
real experimental background~\cite{note}. The combined data are passed
through the same Netscan 
reconstruction and selection programs as used for the real data. The average 
detection efficiencies for NC and CC sample are evaluated to be
(12.8 $\pm$ 0.6)\% and (8.5 $\pm$ 0.6)\%, respectively. 
%different samples are shown in
%table~\ref{tab:eff}.

The ratio of the reconstruction and location efficiency of the  NC $c\bar{c}$  events to that of the whole sample 
of NC events   is found to be 0.496 $\pm$ 0.022. 
This ratio is 0.506 $\pm$ 0.018 in the CC channel.

The lower reconstruction and location efficiency of $c\bar{c}$ 
events is due to the larger hadronic activity in the electronic detector
for these events.

\begin{table} [thc]
\begin{center}
\caption {Background sources and the corresponding event yields in
the $1\mu$ sample.}
\label{tab:bckCC}
\vspace{0.5\baselineskip}
\begin{tabular}{lcc}
{\bf Primary interaction} & {\bf secondary vertex} & {\bf event yield} \\   
\hline
$\nu_{\mu} N \rightarrow \mu^{-} c ~\Sigma^{\pm} X$& $\Sigma^{\pm}
\rightarrow$ 1 prong &  0.002 $\pm$ 0.001 \\
$\nu_{\mu} N \rightarrow \mu^{-} c\ h^{\pm} X$ & $h^{\pm}$ white C1 &
$0.08\pm0.04$ \\
$\nu_{\mu} N \rightarrow \mu^{-} c\ h^{\pm} X$ & $h^{\pm}$ white C3 &
0.06 $\pm$ 0.04 \\
$\nu_{\mu} N \rightarrow \mu^{-} c\ h^{\pm} X$ & $h^{\pm}$ white C5 &
$0.024\pm0.009$ \\
$\nu_{\mu} N \rightarrow \mu^{-} c\ h^{0} X$ & $h^0$ white V2 &
0.01 $\pm$ 0.01 \\
$\nu_{\mu} N \rightarrow \mu^{-} c\ h^{0} X$ & $h^0$ white V4 &
0.006 $\pm$ 0.006 \\
\hline
Total & & 0.18 $\pm$ 0.06 \\
\end{tabular}
\end{center}
\end{table}

\begin{table} [thc]
\begin{center}
\caption {Background sources and the corresponding event yields in
the $0\mu$ sample where $\mu^{\pm}$($e^{\pm}$)  is misidentified.}
\label{tab:bckNC}
\vspace{0.5\baselineskip}
\begin{tabular}{lcc}
{\bf Primary interaction} & {\bf secondary vertex} & {\bf event yield} \\
\hline
$\nu N \rightarrow \mu^{\pm}(\mathrm{e^{\pm}}) c~\Sigma^{\pm} X$& $\Sigma^{\pm}
\rightarrow$ 1 prong &  0.0003 $\pm$ 0.0001 \\
$\nu N \rightarrow  \mu^{\pm}(\mathrm{e^{\pm}}) c\ h^{\pm} X$ & $h^{\pm}$ white C1
&0.013 $\pm$ 0.005 \\
$\nu N \rightarrow  \mu^{\pm}(\mathrm{e^{\pm}}) c\ h^{\pm} X$ & $h^{\pm}$ white
C3 &0.011 $\pm$ 0.005 \\
$\nu N \rightarrow  \mu^{\pm}(\mathrm{e^{\pm}}) c\ h^{\pm} X$ & $h^{\pm}$ white C5 &
0.004 $\pm$ 0.001 \\
$\nu N \rightarrow  \mu^{\pm}(\mathrm{e^{\pm}}) c\ h^{0} X$ & $h^0$ white V2 &
0.001 $\pm$ 0.001 \\
$\nu N \rightarrow  \mu^{\pm}(\mathrm{e^{\pm}}) c\ h^{0} X$ & $h^0$ white V4 &
0.0006 $\pm$ 0.0006\\
$\nu N \rightarrow  \mu^{\pm}(\mathrm{e^{\pm}}) c~\bar{c} X$ & $c\bar{c}\rightarrow $ any 
prong & 0.146 $\pm$ 0.051 \\
\hline
Total & & 0.18 $\pm$ 0.05 \\
\end{tabular}
\end{center}
\end{table}

\section{Background evaluation}
The search is essentially topological even if there are additional
measurements of the momentum and acoplanarity angle to further
validate the candidates. In Table~\ref{tab:bckCC} we list all 
relevant background sources for double charm production in CC
interactions.

A double charm event is topologically indistinguishable from a single
charm event with a primary non-charmed hadron which undergoes either a
decay or an interaction without any visible recoil of the
nuclei. Among the strange hadrons produced in a single charm event,
the $\Sigma^{\pm}$ has the lowest $c\tau$ (a few \cm) and is
relatively abundant (about 1\%). 
Owing to longer  decay lengths there are  negligible 
contributions from decays of other non-charmed hadrons like kaons and
pions.

Decays of strange neutral particles like $\Lambda$ and $K^{0}_{s}$ do
not give a sizeable contribution owing to the acoplanarity cut. 

The interaction length for white interactions is very large, ranging
from a few
metres for single prong interactions to several hundred metres
for multi-prong interactions.  A simulation of these processes using
FLUKA~\cite{fluka} has been carried out and the interaction lengths
for the different processes have been obtained. 
%A number of
%interactions in emulsion has been collected by CHORUS in studies of
%charm decays. This allows experimental cross-checks of the
%calculations. 
In the course of the CHORUS
analysis, 243~m of charged hadron tracks have been followed and 26  white
 interactions~\cite{oscil} have been collected, which allows a cross-check of the
simulation at the 35\% level.
An overall background of 0.18 $\pm$ 0.06 events is
expected in CC interactions, mainly from white interactions.

In Table~\ref{tab:bckNC} 
are listed the background sources and the corresponding event yield in the 
$0\mu$ sample. An overall background of 0.18 $\pm$ 0.05 events is estimated. 
The main background source for the double charm production in NC interaction comes from
charged lepton misidentification in CC double charm events. This background is estimated by scaling the candidate
event in CC interactions with the CC contamination in 0mu sample which is estimated to be about 14\%.

%This value is much lower than for the $1\mu$ sample since the probability 
%for the primary muon to remain unidentified is small.

\section{Results and conclusions}

In summary,  in a sample of 99,245 $1\mu$ events, one  event has decay 
topologies that meet the requirements for associated charm production.
The total background  in the CC sample is estimated to be 0.18 $\pm$ 0.06
events. In the sample of 26,621 $0\mu$ 
events, three events showed decay topologies that are consistent with
associated charm production. The total background in the NC  
sample is estimated to be 0.18 $\pm$ 0.05.

In order to estimate the associated charm production rate in $\nu_{\mu}$
interactions, an additional weight factor needs to be applied to  
$1\mu$ events with $p_{\mu}> 30$ \GeVc, since a small fraction of this
category was not located and analysed. This  
factor was  evaluated from the measured ratio  0.305 $\pm$ 0.002 of $1\mu$
events with $p_{\mu}> 30$ \GeVc to those with $p_{\mu}< 30$ \GeVc; it
was  
 found to be 1.021. In order to evaluate the real number of
$\nu$ NC we have used the measured fraction  of   0.274 $\pm$ 0.005 between
deep-inelastic NC  and CC  $\nu_{\mu}$ interactions. 
This value is
smaller than what was measured by CHARM, CDHS and CCFR collaborations
~\cite{ccfr} since a correction for the non-isoscalarity of the emulsion
target is applied. In NC interactions identification of the neutrino flavor
is not possible in the experiment. Therefore, the NC sample contains a small fraction of $\nu_{e}$,  $\bar\nu_{e}$ and
$\bar\nu_{\mu}$ NC interactions. The contribution of neutrino flavors other than  $\nu_{\mu}$ to the normalization is 
estimated  to be 3.3$\%$. Since in CC interactions  at least one  reconstructed muon in the
spectrometer is required both NC and CC  interactions have
a similar energy threshold which is about  6 \GeV.
 
The relative rate of NC associated charm production is given by
\[
 \frac{\sigma (c\bar{c}\nu)}{\sigma_\mathrm{NC}^\mathrm{DIS}}=
 (\frac{N_\mathrm{obs}^{c\bar{c}}-N_\mathrm{bgr}}{R_\mathrm{\frac{NC}{CC}} 
N_{CC}})\frac{1}{r_\mathrm{loc}}\frac{1}{\epsilon_\mathrm{net}}.  
\]
where:
\begin{itemize}
\item $N_\mathrm{obs}^{c\bar{c}}$ = 3 is the number of candidate events in
      the $0\mu$ sample;
\item $N_\mathrm{bgr}$ = 0.18 $\pm$ 0.05 is the total background in NC sample;
\item $R_\mathrm{\frac{NC}{CC}}$ = 0.285 $\pm$ 0.005 is the effective ratio  between                          
deep-inelastic  NC and  CC interactions;
\item $N_\mathrm{CC}$ = 101,329 is the number of  CC $\nu_{\mu}$ interactions;
\item $\epsilon_\mathrm{net}$ = 0.128 $\pm$ 0.006 is  the average
detection efficiency for the NC $c\bar{c}$ sample.
\item $ r_\mathrm{loc}$ = 0.211 $\pm$ 0.004 is the ratio of the
      reconstruction and location efficiency of 
events with $c\bar{c}$ in NC events  to that of all  CC events.
\end{itemize}
The value obtained for this ratio normalizing to the total neutrino flux with 27 \GeV average neutrino energy is
\[
 \frac{\sigma
 (c\bar{c}\nu)}{\sigma_\mathrm{NC}^\mathrm{DIS}}=(3.62^{+2.95}_{-2.42}(\mbox{stat})\pm
 0.54(\mbox{syst}))\times 10^{-3}. 
\]
The statistical error is derived using a 68$\%$ confidence interval in the unified approach for the analysis of small 
signals in 
the presence of background~\cite{feldman}.  We have accounted for a
systematic uncertainty of 15\% coming from the 
efficiency estimation by Monte Carlo modeling. The energy threshold of associated charm  production is
significantly higher than for NC interactions. Based on the HERWIG event generator, an energy 
threshold cut of 35 \GeV is applied to NC interactions.
The relative rate of NC associated charm production is also evaluated with the 
energy threshold cut and found to be $(7.30^{+5.95}_{-4.88}(\mbox{stat})\pm
 1.09(\mbox{syst}))\times 10^{-3}$. The average neutrino energy above this threshold is 73 \GeV, to be
compared with the average visible energy of our candidates of 44 \GeV.
The neutrino spectrum for NC interactions is peaked at 
about 25 \GeV and decreases very rapidly. Therefore, we expect events to occur close the threshold.
The result is consistent with
the E531~\cite{e531} and NuTeV~\cite{nutev} measurements. The measured cross-section is consistent with our measurement on NC production of  
$\mathrm{J/\Psi}$~\cite{choruspsi}\footnote{We assume that $\mathrm{J/\Psi}$ production is 1/6 of the total ccbar production}.
Given the fact that the cross-section is predicted to have a strong energy dependence and taking into account that the neutrino energy spectrum of the beam used in this
experiment peaks at values where this cross-section is predicted to be low, our result is compatible with the  prediction of the $\mathrm{Z^0}$-gluon fusion 
model~\cite{barger}. 
In the framework of this model, one obtains the relative rate of NC associated charm production 
as $\sim 4\times 10^{-3}$.

With the observation of one event for associated charm production in CC interactions, 
we obtain for the relative rate an upper limit at 90\% C.L.~\cite{feldman} of  
\[
 \frac{\sigma (c\bar{c}\mu^{-})}{\sigma_\mathrm{CC}}<9.69 \times 10^{-4},
\]
normalizing to the total neutrino flux.
Due to different  energy
thresholds in single and associated charm production,   the production cross-section
relative to CC interactions is estimated with the energy cut of 35 \GeV.
An upper limit for the relative rate with the energy threshold cut is found to be $2.24  
\times 10^{-3}$. The average neutrino energy above this threshold is 73 \GeV.
Since the topology of the single candidate event is not one of the most likely 
background channels, it is justified to give a cross-section for this process.
Based on the
single event,  the production cross-section relative to CC interactions is 
$1.95^{+3.22}_{-1.44}(\mbox{stat})\pm 0.29(\mbox{syst})\times 10^{-4}$ with a systematic 
error of 15\% from the flux normalization.  
The rate of CC  associated charm production with the threshold cut is obtained to be 
$(4.50^{+7.44}_{-3.33}(\mbox{stat})\pm 0.68(\mbox{syst}))\times 10^{-4}$.
The cross-section predicted by the QCD inspired
parton model~\cite{Hagiwara:1980nu} has a strong energy dependence. Although  the cross-section of this process at the average  energy of the CHORUS $\nu_\mu$ beam
is low, the measured production rate   is  in agreement   with the prediction of  the QCD inspired 
parton model. The relative 
rate of CC associated charm production is calculated to be $\sim 2\times 10^{-4}$ within the  framework of this model.
The {\em a posteriori} probability that the background for the topology with two neutral decays gives one event is 0.016.  Taking into
account that this was not the only topology searched for, it is difficult to convert this into a uniquely defined 
confidence level.
However, given the special topology, it is very likely that this event constitutes an observation of associated 
charm production in CC
interactions.

\section{Acknowledgments}  

We gratefully acknowledge the help and support of our numerous
technical collaborators who contributed to the detector construction
and operation. We thank the neutrino beam staff for the competent
assistance ensuring the excellent performance of the facility.  The
accumulation of a large data sample in this experiment has also been
made possible thanks to the efforts of the crew operating the CERN PS
and SPS. The general technical support from EP (ECP) and IT Divisions
is gratefully acknowledged.

The experiment has been made possible by grants from our funding
agencies: the Institut Interuniversitaire des Sciences
Nucl$\acute{\mathrm{e}}$aires and the Interuniversitair Instituut voor
Kernwetenschappen (Belgium), The Israel Science foundation (Grant
328/94) and the Technion Vice President Fund for the Promotion of
Research (Israel), CERN (Geneva, Switzerland), the German
Bundesministerium f$\ddot{\mathrm{u}}$r Bildung und Forschung (Grant
057MS12P(0)) (Germany), the Institute of Theoretical and Experimental
Physics (Moscow, Russia), the Instituto Nazionale di Fisica Nucleare
(Italy), the Promotion and Mutual Aid Corporation for Private Schools
of Japan and Japan Society for the Promotion of Science (Japan), the
Korea Research Foundation Grant (KRF-99-005-D00004) (Republic of
Korea), the Foundation for Fundamental Research on Matter FOM and the
National Scientific Research Organization NWO (The Netherlands) and
the Scientific and Technical Research Council of Turkey (Turkey).

\end{document}